\documentclass[pdflatex,sn-mathphys-num]{sn-jnl}%
\usepackage{lineno}
\newgeometry{           
    twoside=false,
    top=3.5cm,
    left=4.7cm,
    right=4.7cm,
    bottom=4.5cm
}
\usepackage{array}
\usepackage[table]{xcolor}
\usepackage[most]{tcolorbox}
\newcommand\blfootnote[1]{%
  \begingroup
  \renewcommand\thefootnote{}\footnote{\hskip-8.75pt #1}%
  \addtocounter{footnote}{-1}%
  \endgroup
}
\usepackage[backend=biber,style=nature]{biblatex}
\usepackage{xurl}
\hypersetup{breaklinks=true}
\renewcommand{\mkbibbrackets}[1]{\mkbibparens{#1}}
\AtEveryBibitem{\clearfield{issn}}
\AtEveryBibitem{\clearfield{month}}
\AtEveryBibitem{\clearfield{day}}
\AtEveryBibitem{\clearlist{language}}
\newcommand\Tstrut{\rule[-2mm]{0pt}{7mm}}
\definecolor{background}{HTML}{EEECE1}
\usepackage{titlesec}
\titleformat{name=\section}
{\filcenter\Large\bfseries}
{}
{0em}
{}
\titleformat{name=\subsection}
{\large\bfseries}
{}
{0em}
{}
\titleformat{name=\subsubsection}
{\bfseries}
{}
{0em}
{}

\title{\mbox{Lessons from complexity theory} \mbox{for AI governance}} 
\author[1,2]{\fnm{Noam} \sur{Kolt}}\email{noam.kolt@mail.huji.ac.il}
 
\author[1]{\fnm{Michal} \sur{Shur-Ofry}}

\author[3]{\fnm{Reuven} \sur{Cohen}}

\affil[1]{\orgdiv{Faculty of Law}, \orgname{Hebrew University}%
}

\affil[2]{\orgdiv{School of Computer Science and Engineering}, \orgname{Hebrew University}%
}

\affil[3]{\orgdiv{Department of Mathematics}, \orgname{Bar-Ilan University}%
}

\date{}

\abstract{%
The study of complex adaptive systems, pioneered in physics, biology, and the social sciences, offers important lessons for AI governance. Contemporary AI systems and the environments in which they operate exhibit many of the properties characteristic of complex systems, including nonlinear growth patterns, emergent phenomena, and cascading effects that can lead to tail risks. Complexity theory can help illuminate the features of AI that pose central challenges for policymakers, such as feedback loops induced by training AI models on synthetic data and the interconnectedness between AI systems and critical infrastructure. Drawing on insights from other domains shaped by complex systems, including public health and climate change, we examine how efforts to govern AI are marked by deep uncertainty. To contend with this challenge, we propose a set of complexity-compatible principles concerning the timing and structure of AI governance, and the risk thresholds that should trigger regulatory intervention.%
}


\begin{document}
\maketitle

\blfootnote{Correspondence to: \href{mailto:noam.kolt@mail.huji.ac.il}{noam.kolt@mail.huji.ac.il}. 
}

\newpage

\section{Introduction}\label{introduction}

Discussion of the impact of AI and approaches to governing the technology have become increasingly polarized. Scholars and practitioners concerned about the risks from AI systems fiercely debate the appropriate goals, scope, and timing of regulatory policy and intervention \cite{Price_Connelly_2023, noauthor_2023b}, often divided along disciplinary lines or research communities \cite{Lazar_Nelson_2023}. The discourse is to a large extent influenced by conceptual framing. Some characterize AI as a highly consequential software product or service \cite{Bommasani_Kapoor_Klyman_Longpre_Ramaswami_Zhang_Schaake_Ho_Narayanan_Liang_2024,Kapoor_Bommasani_Klyman_Longpre_Ramaswami_Cihon_Hopkins_Bankston_Biderman_Bogen_Chowdhury_Engler_Henderson_Jernite_Lazar_Maffulli_Nelson_Pineau_Skowron_Song_Storchan_Zhang_Ho_Liang_Narayanan_2024}, while others characterize AI as a societal-scale transformation that presents unprecedented risks \cite{Anwar_Saparov_Rando_Paleka_Turpin_Hase_Lubana_Jenner_Casper_Sourbut_Edelman_Zhang_Günther_Korinek_Hernandez-Orallo_Hammond_Bigelow_Pan_Langosco_Korbak_Zhang_Zhong_hEigeartaigh_Recchia_Corsi_Chan_Anderljung_Edwards_Petrov_Witt_Motwani_Bengio_Chen_Torr_Albanie_Maharaj_Foerster_Tramèr_He_Kasirzadeh_Choi_Krueger_2024,Bengio_Hinton_Yao_Song_Abbeel_Darrell_Harari_Zhang_Xue_Shalev-Shwartz_Hadfield_Clune_Maharaj_Hutter_Baydin_McIlraith_Gao_Acharya_Krueger_Dragan_Torr_Russell_Kahneman_Brauner_Mindermann_2024}.

We propose a different lens, grounded in decades of interdisciplinary research in physics, biology, and the social sciences: \emph{analyzing AI systems, their development process, and the environments in which they operate as complex systems}. Complex systems are systems comprised of multiple interacting components. Examples of such systems, in the natural and human world, include insect colonies, urban environments, social networks, and financial markets \cite{Cohen_Havlin_2010,Miller_Page_2007,Mitchell_2009}.

Complexity theory demonstrates that complex systems of different kinds share common traits, including the \emph{emergence} of system-level properties and patterns despite the absence of central control or design, and \emph{nonlinear dynamics} that defy simple cause-effect relations. Small changes in a complex system's topology and interactions among its components may result in very different overall effects. Complex systems thus entail inherent \emph{unpredictability} and are susceptible to rare but substantial \emph{cascades} and the materialization of \emph{tail risks} with potentially far-reaching consequences \cite{Bashan_Berezin_Buldyrev_Havlin_2013,Buldyrev_Parshani_Paul_Stanley_Havlin_2010,Cohen_Havlin_2010,Gao_Bashan_Shekhtman_Havlin_2022,Li_Bashan_Buldyrev_Stanley_Havlin_2012,Yang_Nishikawa_Motter_2017}.

\begin{tcolorbox}[breakable,enhanced jigsaw, pad at break*=1mm,colbacktitle=background,colback=background,coltitle=black,toptitle=3mm,bottomtitle=3mm,width=\linewidth,fonttitle=\bfseries,parbox=true,title=Box 1: Defining complex systems,attach boxed title to top,phantom={\phantomsection\hypertarget{box1}},boxed title style={colframe=white,boxrule=1pt,frame hidden},boxsep=3pt,left=6pt,right=6pt,top=6pt,bottom=6pt]
\small

``a complex system\ldots{[}is{]}\ldots{} one made up of a large number of parts that interact in a non-simple way. In such systems, the whole is more than the sum of its parts \ldots{} in the important pragmatic sense that, given the properties of the parts and the laws of their interaction, it is not a trivial matter to infer the properties of the whole.'' \hfill--- Herbert A. Simon, \emph{The Architecture of Complexity} (1962) \cite{Simon_1962}
\end{tcolorbox}

Methodologies developed in complexity theory enable researchers to better understand complex systems and, where appropriate, design policies to address the associated societal challenges. Drawing on multiple studies that suggest that central aspects of contemporary AI systems bear the hallmarks of complexity \parencite{Dobbe_2022,Hendrycks_2025,Holtzman_West_Zettlemoyer_2023,Leveson_2012,Macrae_2022,Nanda_Chan_Lieberum_Smith_Steinhardt_2022,Power_Burda_Edwards_Babuschkin_Misra_2022,Rakova_Dobbe_2023,Rismani_Shelby_Smart_Jatho_Kroll_Moon_Rostamzadeh_2023,Shelby_Rismani_Henne_Moon_Rostamzadeh_Nicholas_Yilla-Akbari_Gallegos_Smart_Garcia_Virk_2023,Wei_Tay_Bommasani_Raffel_Zoph_Borgeaud_Yogatama_Bosma_Zhou_Metzler_Chi_Hashimoto_Vinyals_Liang_Dean_Fedus_2022,Weidinger_Rauh_Marchal_Manzini_Hendricks_Mateos-Garcia_Bergman_Kay_Griffin_Bariach_Gabriel_Rieser_Isaac_2023,Yoo_2024}, we make three primary contributions: \emph{\textbf{First}}, we unpack the characterization of AI systems as complex systems. \emph{\textbf{Second}}, we explore the implications of this characterization for the challenges involved in governing AI. \emph{\textbf{Third}}, drawing on insights from complexity and other domains shaped by complex systems, we propose a series of complexity-compatible principles to assist policymakers in developing more effective mechanisms for governing AI.

\section{AI and complexity}\label{ai-and-complexity}

Our characterization of AI systems as complex systems focuses on several properties increasingly identified in AI systems, the processes through which they are trained, and the environments in which they are deployed. The properties we focus on are \emph{nonlinear growth}, \emph{unpredictable scaling and emergence}, \emph{feedback loops}, \emph{cascading effects}, and \emph{tail risks.} While the list is not exhaustive and varies substantially across different forms of AI and application domains, these properties illuminate some of the distinctive governance challenges posed by AI.

\subsection{Nonlinear growth}\label{nonlinear-growth}

In recent years, there has been an exponential increase in many of the key inputs into AI development. The computational resources for training AI models have, on average, grown by a factor of four to five each year between 2010 and 2024 \cite{Sevilla_2024}. The size of datasets used for training has also increased significantly. For example, language training dataset size has increased by a factor of three each year in recent years \cite{EpochAI_2023}. Meanwhile, efficiency of computation has continued to improve exponentially \cite{Pilz_Heim_Brown_2024,Ho_Besiroglu_Erdil_Owen_Rahman_Guo_Atkinson_Thompson_Sevilla_2024}, alongside corporate investment that has increased by more than an order of magnitude \cite{noauthor_nodate}.

During this period, the capabilities of AI systems have improved dramatically. AI systems can now outperform humans on some tasks \cite{Morris_Sohl-Dickstein_Fiedel_Warkentin_Dafoe_Faust_Farabet_Legg_2024}, including certain tasks relating to visual question answering, natural language understanding, and text annotation \cite{Gilardi_Alizadeh_Kubli_2023, noauthor_nodate}. Several benchmarks previously used to evaluate AI systems have, due to improvements in the capabilities of AI, been rendered obsolete \cite{Deng_Dong_Socher_Li_Li_Fei-Fei_2009,Wang_Pruksachatkun_Nangia_Singh_Michael_Hill_Levy_Bowman_2019}. AI systems have also achieved superhuman feats in various scientific fields, including biology \cite{Abramson_Adler_Dunger2024,Jumper_Evans_Pritzel_Green_Figurnov_Ronneberger_Tunyasuvunakool_Bates_Žídek_Potapenko_Bridgland_Meyer_Kohl_Ballard_Cowie_Romera-Paredes_Nikolov_Jain_Adler_Back_Petersen_Reiman_Clancy_Zielinski_Steinegger_Pacholska_Berghammer_Bodenstein_Silver_Vinyals_Senior_Kavukcuoglu_Kohli_Hassabis_2021}, mathematics \cite{Romera-Paredes_Barekatain_Novikov_Balog_Kumar_Dupont_Ruiz_Ellenberg_Wang_Fawzi_Kohli_Fawzi_2024}, weather forecasting \cite{Lam_Sanchez-Gonzalez_Willson_Wirnsberger_Fortunato_Alet_Ravuri_Ewalds_Eaton-Rosen_Hu_Merose_Hoyer_Holland_Vinyals_Stott_Pritzel_Mohamed_Battaglia_2023}, and materials science \cite{Merchant_Batzner_Schoenholz_Aykol_Cheon_Cubuk_2023}. Importantly, as we illustrate, these improvements in performance may themselves exhibit properties of complexity.

\subsection{Scaling, emergence, and unpredictability}\label{scaling-emergence-and-unpredictability}

The effect of increases in the inputs into widely used AI systems, especially foundation models, on the performance of those systems resembles patterns characteristic of other complex systems. This phenomenon has been observed both in \textit{model training}, since the advent of foundation models \cite{Bommasani_Hudson_Adeli_Altman_Arora_Arx_Bernstein_Bohg_Bosselut_Brunskill_Brynjolfsson_Buch_Card_Castellon_Chatterji_Chen_Creel_Davis_Demszky_Donahue_Doumbouya_Durmus_Ermon_Etchemendy_Ethayarajh_Fei-Fei_Finn_Gale_Gillespie_Goel_Goodman_Grossman_Guha_Hashimoto_Henderson_Hewitt_Ho_Hong_Hsu_Huang_Icard_Jain_Jurafsky_Kalluri_Karamcheti_Keeling_Khani_Khattab_Koh_Krass_Krishna_Kuditipudi_Kumar_Ladhak_Lee_Lee_Leskovec_Levent_Li_Li_Ma_Malik_Manning_Mirchandani_Mitchell_Munyikwa_Nair_Narayan_Narayanan_Newman_Nie_Niebles_Nilforoshan_Nyarko_Ogut_Orr_Papadimitriou_Park_Piech_Portelance_Potts_Raghunathan_Reich_Ren_Rong_Roohani_Ruiz_Ryan_Ré_Sadigh_Sagawa_Santhanam_Shih_Srinivasan_Tamkin_Taori_Thomas_Tramèr_Wang_Wang_Wu_Wu_Wu_Xie_Yasunaga_You_Zaharia_Zhang_Zhang_Zhang_Zhang_Zheng_Zhou_Liang_2022}, and in \textit{model inference}, following the development of reasoning models (i.e., models that ``think'' using chain-of-thought at run-time) \cite{xu2025towards}.

In model training, cross entropy loss---the main metric used to measure training performance---has been shown to scale (decrease) in a power-law relationship with model size, dataset size, and the amount of compute used in training \cite{Bahri_Dyer_Kaplan_Lee_Sharma_2024,Henighan_Kaplan_Katz_Chen_Hesse_Jackson_Jun_Brown_Dhariwal_Gray_Hallacy_Mann_Radford_Ramesh_Ryder_Ziegler_Schulman_Amodei_McCandlish_2020,Hoffmann_Borgeaud_Mensch_Buchatskaya_Cai_Rutherford_deLasCasas_Hendricks_Welbl_Clark_Hennigan_Noland_Millican_vandenDriessche_Damoc_Guy_Osindero_Simonyan_Elsen_Vinyals_Rae_Sifre_2022,Kaplan_McCandlish_Henighan_Brown_Chess_Child_Gray_Radford_Wu_Amodei_2020}. These ``scaling laws'' suggest that the performance of AI models (measured by cross entropy loss) may continue to improve with increases in the inputs used in training. However, the \emph{specific capabilities} acquired by these models in practice, that is, their ability to perform particular real-world tasks, remains highly unpredictable and can appear to emerge suddenly \cite{Power_Burda_Edwards_Babuschkin_Misra_2022, Ganguli_Hernandez_Lovitt_Askell_Bai_Chen_Conerly_Dassarma_Drain_Elhage_ElShowk_Fort_Hatfield-Dodds_Henighan_Johnston_Jones_Joseph_Kernian_Kravec_Mann_Nanda_Ndousse_Olsson_Amodei_Brown_Kaplan_McCandlish_Olah_Amodei_Clark_2022,Ruan_Maddison_Hashimoto_2024,Schaeffer_Schoelkopf_Miranda_Mukobi_Madan_Ibrahim_Bradley_Biderman_Koyejo_2024}. An early illustration of this phenomenon was observed in 2020 with GPT-3 which, although structurally similar to prior models, due to its larger size gained the qualitatively new ability to learn to perform new tasks after being provided a few demonstrations of those tasks (known as ``few-shot learning'') \cite{Brown_Mann_Ryder_Subbiah_Kaplan_Dhariwal_Neelakantan_Shyam_Sastry_Askell_Agarwal_Herbert-Voss_Krueger_Henighan_Child_Ramesh_Ziegler_Wu_Winter_Hesse_Chen_Sigler_Litwin_Gray_Chess_Clark_Berner_McCandlish_Radford_Sutskever_Amodei_2020}.

More recently, progress in the development of reasoning models---such as OpenAI's o series of models \cite{OpenAI_2024} and DeepSeek's R series of models \cite{guo2025deepseek} released in 2024 and 2025, respectively---demonstrates an equivalent phenomenon. Increasing the scale of inference (test-time) compute has resulted in systems gaining qualitatively new abilities, including exceeding human PhD-level accuracy on certain STEM-related benchmarks \cite{Brown_Juravsky_Ehrlich_Clark_Le_Ré_Mirhoseini_2024,Snell_Lee_Xu_Kumar_2024,Stroebl_Kapoor_Narayanan_2024,Wu_Sun_Li_Welleck_Yang_2024, li2025misfitting}.

Seen through the lens of complexity theory, the unpredictable emergence of new capabilities in AI models can be analogized to phase transitions in physical and biological systems \cite{Anderson_1972,Stanley_1987}, such as water freezing or boiling when it reaches a certain temperature, or the emergence of cognition from multiple neural interactions \cite{Lubana_Kawaguchi_Dick_Tanaka_2024,Pan_Bhatia_Steinhardt_2021}. Similarly, AI systems appear to acquire new, qualitatively different abilities when a certain threshold is reached, either in training or at inference. The exact scope and nature of new AI abilities, however, is unpredictable. For instance, OpenAI's o3 model was able to solve over 25\% of the problems in the FrontierMath benchmark, surpassing the 2\% achieved by previous models and defying Terence Tao's prediction that the problems would ``resist AIs for several years at least'' \cite{EpochAI_2024}.

\subsection{Feedback loops}\label{feedback-loops}

Like other complex systems, AI systems interact with their environments and are prone to feedback loops that can generate self-reinforcing processes. These can occur, for instance, where the output of an AI model influences human behavior that is then incorporated into the data used to refine the model or train future models. For example, various algorithms used to predict housing prices can influence real-world housing prices, which then influence future price predictions, and so on \cite{Fu_Jin_Liu_2022}. This kind of feedback loop in which predictions that support decisions influence the very outcomes they aim to predict is known as ``performative prediction'' \cite{Hardt_Mendler-Dünner_2023,Healy_2015,Perdomo_Zrnic_Mendler-Dünner_Hardt_2020}. Feedback loops also commonly arise in the context of content recommendation. Recommender systems respond to users' selection of content by recommending similar content which, in turn, reinforces users' existing content preferences \cite{Cen_Ilyas_Allen_Li_Madry_2024,Pedreschi_Pappalardo_Ferragina_Baeza-Yates_Barabási_Dignum_Dignum_Eliassi-Rad_Giannotti_Kertész_Knott_Ioannidis_Lukowicz_Passarella_Pentland_Shawe-Taylor_Vespignani_2025,Stray_Halevy_Assar_Hadfield-Menell_Boutilier_Ashar_Bakalar_Beattie_Ekstrand_Leibowicz_MoonSehat_Johansen_Kerlin_Vickrey_Singh_Vrijenhoek_Zhang_Andrus_Helberger_Proutskova_Mitra_Vasan_2024,Williams_Carroll_Narang_Weisser_Murphy_Dragan_2024}.

The widespread use of foundation models exacerbates such feedback loops. Because the outputs of foundation models are increasingly incorporated into publicly available data repositories, which are then included in the training data of future models, errors and biases in earlier models could compound with each successive generation of models \cite{Bender_Gebru_McMillan-Major_Shmitchell_2021,Shur-Ofry_2025,Taori_Hashimoto_2023}. For example, anti-consumer biases in language models used to perform legal tasks could intensify if the biased outputs of those models are used to train future models \cite{Kolt_2022}.

Feedback loops might also ensue as humans tasked with annotating data for training AI models outsource their work to other AI models \cite{Veselovsky_Ribeiro_Cozzolino_Gordon_Rothschild_West_2023,Veselovsky_Ribeiro_West_2023,Wu_Zhu_Albayrak_Axon_Bertsch_Deng_Ding_Guo_Gururaja_Kuo_Liang_Liu_Mandal_Milbauer_Ni_Padmanabhan_Ramkumar_Sudjianto_Taylor_Tseng_Vaidos_Wu_Wu_Yang_2023}, or are influenced by their use of AI models \cite{Jakesch_Bhat_Buschek_Zalmanson_Naaman_2023}. In addition, training models on large quantities of synthetic data (i.e., data generated by other AI models) \cite{Fan_Chen_Krishnan_Katabi_Isola_Tian_2024,Lee_Phatale_Mansoor_Mesnard_Ferret_Lu_Bishop_Hall_Carbune_Rastogi_Prakash_2024,Liu_Wei_Liu_Si_Zhang_Rao_Zheng_Peng_Yang_Zhou_Dai_2024,Singh_Co-Reyes_Agarwal_Anand_Patil_Garcia_Liu_Harrison_Lee_Xu_Parisi_Kumar_Alemi_Rizkowsky_Nova_Adlam_Bohnet_Elsayed_Sedghi_Mordatch_Simpson_Gur_Snoek_Pennington_Hron_Kenealy_Swersky_Mahajan_Culp_Xiao_Bileschi_Constant_Novak_Liu_Warkentin_Bansal_Dyer_Neyshabur_Sohl-Dickstein_Fiedel_2024} can in some circumstances degrade the quality of the resulting models \cite{Alemohammad_Casco-Rodriguez_Luzi_Humayun_Babaei_LeJeune_Siahkoohi_Baraniuk_2023,Bohacek_Farid_2023,Gerstgrasser_Schaeffer_Dey_Rafailov_Korbak_Sleight_Agrawal_Hughes_Pai_Gromov_Roberts_Yang_Donoho_Koyejo_2024,Kazdan_Schaeffer_Dey_Gerstgrasser_Rafailov_Donoho_Koyejo_2024,Shumailov_Shumaylov_Zhao_Gal_Papernot_Anderson_2024,Shumailov_Shumaylov_Zhao_Papernot_Anderson_Gal_2024}.\footnote{Synthetic data has, however, been central to the development of reasoning models such as DeepSeek R1 \cite{guo2025deepseek}.}  This may be exacerbated by the fact that detecting AI-generated content (e.g., by using watermarks) \cite{Kirchenbauer_Geiping_Wen_Katz_Miers_Goldstein_2023} and excluding it from training datasets remains difficult \cite{Kirchenbauer_Geiping_Wen_Shu_Saifullah_Kong_Fernando_Saha_Goldblum_Goldstein_2023,Sadasivan_Kumar_Balasubramanian_Wang_Feizi_2024}. Studies suggest that a growing fraction of content on the internet is already dominated by synthetic content, including synthetic content produced by models that are themselves trained on synthetic content \cite{Huang_Siddarth_2023,Liang_Izzo_Zhang_Lepp_Cao_Zhao_Chen_Ye_Liu_Huang_McFarland_Zou_2024,Thompson_Dhaliwal_Frisch_Domhan_Federico_2024}. 

Novel feedback loops could also arise as AI models are increasingly used to evaluate the safety of other AI models \cite{Balloccu_Schmidtová_Lango_Dusek_2024,Panickssery_Bowman_Feng_2024,Perez_Huang_Song_Cai_Ring_Aslanides_Glaese_McAleese_Irving_2022,Perez_Ringer_Lukosiute_Nguyen_Chen_Heiner_Pettit_Olsson_Kundu_Kadavath_Jones_Chen_Mann_Israel_Seethor_McKinnon_Olah_Yan_Amodei_Amodei_Drain_Li_Tran-Johnson_Khundadze_Kernion_Landis_Kerr_Mueller_Hyun_Landau_Ndousse_Goldberg_Lovitt_Lucas_Sellitto_Zhang_Kingsland_Elhage_Joseph_Mercado_DasSarma_Rausch_Larson_McCandlish_Johnston_Kravec_ElShowk_Lanham_Telleen-Lawton_Brown_Henighan_Hume_Bai_Hatfield-Dodds_Clark_Bowman_Askell_Grosse_Hernandez_Ganguli_Hubinger_Schiefer_Kaplan_2023} or assist in conducting AI safety research \cite{Bowman_Hyun_Perez_Chen_Pettit_Heiner_Lukošiūtė_Askell_Jones_Chen_Goldie_Mirhoseini_McKinnon_Olah_Amodei_Amodei_Drain_Li_Tran-Johnson_Kernion_Kerr_Mueller_Ladish_Landau_Ndousse_Lovitt_Elhage_Schiefer_Joseph_Mercado_DasSarma_Larson_McCandlish_Kundu_Johnston_Kravec_Showk_Fort_Telleen-Lawton_Brown_Henighan_Hume_Bai_Hatfield-Dodds_Mann_Kaplan_2022,Burns_Izmailov_Kirchner_Baker_Gao_Aschenbrenner_Chen_Ecoffet_Joglekar_Leike_Sutskever_Wu_2024}. While forecasting the precise contours of these interactions will likely be impossible, suffice to say that research in complexity theory suggests these phenomena could lead to rapid and potentially dangerous self-reinforcing processes, especially in the case of interconnected systems and networks \cite{Pan_Jones_Jagadeesan_Steinhardt_2024,Wyllie_Shumailov_Papernot_2024}.

\subsection{Interconnectedness, cascading effects, and tail risks}\label{interconnectedness-cascading-effects-and-tail-risks}

Complexity theory sheds light on the vulnerability of interdependent networks to cascading effects whereby damage to a small number of nodes (i.e., components comprising the system) in one network can have an outsized impact on other interconnected systems, potentially causing large-scale damage \cite{Bashan_Berezin_Buldyrev_Havlin_2013,Buldyrev_Parshani_Paul_Stanley_Havlin_2010,Cohen_Havlin_2010,Gao_Bashan_Shekhtman_Havlin_2022,Li_Bashan_Buldyrev_Stanley_Havlin_2012,Yang_Nishikawa_Motter_2017}. For example, power outages can cause internet outages that then cause further power outages, which can affect additional interconnected networks, such as telecommunications networks \cite{Buldyrev_Parshani_Paul_Stanley_Havlin_2010}. 

Similar dynamics could---and perhaps already do---arise in AI, especially when AI systems are integrated into other systems. The prevailing AI paradigm in which foundation models perform downstream applications across multiple domains is particularly vulnerable to cascading effects. Minor defects in foundation models can propagate across the myriad settings in which they are deployed \cite{Bommasani_Hudson_Adeli_Altman_Arora_Arx_Bernstein_Bohg_Bosselut_Brunskill_Brynjolfsson_Buch_Card_Castellon_Chatterji_Chen_Creel_Davis_Demszky_Donahue_Doumbouya_Durmus_Ermon_Etchemendy_Ethayarajh_Fei-Fei_Finn_Gale_Gillespie_Goel_Goodman_Grossman_Guha_Hashimoto_Henderson_Hewitt_Ho_Hong_Hsu_Huang_Icard_Jain_Jurafsky_Kalluri_Karamcheti_Keeling_Khani_Khattab_Koh_Krass_Krishna_Kuditipudi_Kumar_Ladhak_Lee_Lee_Leskovec_Levent_Li_Li_Ma_Malik_Manning_Mirchandani_Mitchell_Munyikwa_Nair_Narayan_Narayanan_Newman_Nie_Niebles_Nilforoshan_Nyarko_Ogut_Orr_Papadimitriou_Park_Piech_Portelance_Potts_Raghunathan_Reich_Ren_Rong_Roohani_Ruiz_Ryan_Ré_Sadigh_Sagawa_Santhanam_Shih_Srinivasan_Tamkin_Taori_Thomas_Tramèr_Wang_Wang_Wu_Wu_Wu_Xie_Yasunaga_You_Zaharia_Zhang_Zhang_Zhang_Zhang_Zheng_Zhou_Liang_2022,Feng_Park_Liu_Tsvetkov_2023}. While the homogenization introduced by foundation models promotes efficiency (expensive-to-train models can be cheaply reused and adapted to many applications) it also gives rise to new risks familiar to complexity researchers. Safety failures resulting from foundation models might not be independent or isolated from one another, but correlated and connected \cite{Toups_Bommasani_Creel_Bana_Jurafsky_Liang_2024}. For example, systems that exhibit misalignment in one context (e.g., they produce insecure code) tend to exhibit misalignment in other, ostensibly unrelated contexts (e.g., they provide malicious advice and act deceptively) \cite{betley2025emergent}. Meanwhile, vulnerability to a particular type of adversarial attack can diffuse across multiple domains in which a model or agent is deployed or across different models or agents built using similar architecture \cite{Zou_Wang_Carlini_Nasr_Kolter_Fredrikson_2023, li2025commercial}.

Cascading effects could compound as AI systems are integrated into external networks and infrastructure. For example, autonomous agents tasked with pursuing complex goals in safety-critical domains, such as financial markets and essential services, could have highly unpredictable and adverse consequences \cite{Chan_Salganik_Markelius_Pang_Rajkumar_Krasheninnikov_Langosco_He_Duan_Carroll_Lin_Mayhew_Collins_Molamohammadi_Burden_Zhao_Rismani_Voudouris_Bhatt_Weller_Krueger_Maharaj_2023,Cohen_Kolt_Bengio_Hadfield_Russell_2024,Kolt_2024b}. As autonomous agents are increasingly integrated into other systems potential cascading effects could become even broader and harder to predict \cite{Ruan_Dong_Wang_Pitis_Zhou_Ba_Dubois_Maddison_Hashimoto_2024}. A central factor in assessing these effects and associated risks is the level of interconnectedness between AI systems and other systems with which they interact \cite{Shur-Ofry_2024}. Higher levels of interconnectedness imply more vulnerability of risks percolating from an AI system to other systems \cite{hammond2025multi}. Meanwhile, AI systems that operate in closed environments, or in settings with only limited interconnectedness, likely pose less severe risks.

As a result of interconnectedness, feedback loops, and cascading effects, complex systems are particularly susceptible to catastrophic tail risks. For instance, interconnected feedback loops can upend financial markets in unpredictable high-impact events sometimes described as black swans \cite{Taylor_Williams_2009}. AI systems could give rise to similar risks \cite{Anwar_Saparov_Rando_Paleka_Turpin_Hase_Lubana_Jenner_Casper_Sourbut_Edelman_Zhang_Günther_Korinek_Hernandez-Orallo_Hammond_Bigelow_Pan_Langosco_Korbak_Zhang_Zhong_hEigeartaigh_Recchia_Corsi_Chan_Anderljung_Edwards_Petrov_Witt_Motwani_Bengio_Chen_Torr_Albanie_Maharaj_Foerster_Tramèr_He_Kasirzadeh_Choi_Krueger_2024,Bengio_Hinton_Yao_Song_Abbeel_Darrell_Harari_Zhang_Xue_Shalev-Shwartz_Hadfield_Clune_Maharaj_Hutter_Baydin_McIlraith_Gao_Acharya_Krueger_Dragan_Torr_Russell_Kahneman_Brauner_Mindermann_2024}. To illustrate, a malfunctioning AI system used to control wastewater treatment facilities might not only cause direct harm by discharging untreated effluent, but could also have wider adverse effects on human health and marine life \cite{Richards_Tzachor_Avin_Fenner_2023}. These tail risks could become more acute if AI systems are integrated into critical infrastructure. For instance, if AI systems used to control water infrastructure that cools data centers used to train or operate AI systems, then single-system failures---whether resulting from accidental malfunction or malicious adversarial attack---could have far wider consequences \cite{Galaz_Centeno_Callahan_Causevic_Patterson_Brass_Baum_Farber_Fischer_Garcia_McPhearson_Jimenez_King_Larcey_Levy_2021,Richards_Tzachor_Avin_Fenner_2023,Tzachor_Devare_King_Avin_ÓhÉigeartaigh_2022}. 

Importantly, tail risks from AI might not necessarily materialize due solely to the defects in a particular AI system percolating to other systems. Instead, tail risks may arise through the interaction of AI systems with broader sociotechnical structures \cite{Dobbe_2022,Hendrycks_2025,Lazar_Nelson_2023,Leveson_2012,Macrae_2022,Rakova_Dobbe_2023,Rismani_Shelby_Smart_Jatho_Kroll_Moon_Rostamzadeh_2023,Shelby_Rismani_Henne_Moon_Rostamzadeh_Nicholas_Yilla-Akbari_Gallegos_Smart_Garcia_Virk_2023,Weidinger_Rauh_Marchal_Manzini_Hendricks_Mateos-Garcia_Bergman_Kay_Griffin_Bariach_Gabriel_Rieser_Isaac_2023,Yoo_2024}. For example, economic incentives and corporate governance structures may prompt companies to deploy AI systems and enable their use in high-stakes domains without sufficient safeguards \cite{Ganguli_Hernandez_Lovitt_Askell_Bai_Chen_Conerly_Dassarma_Drain_Elhage_ElShowk_Fort_Hatfield-Dodds_Henighan_Johnston_Jones_Joseph_Kernian_Kravec_Mann_Nanda_Ndousse_Olsson_Amodei_Brown_Kaplan_McCandlish_Olah_Amodei_Clark_2022,Askell_Brundage_Hadfield_2019,Kolt_2024a,Dafoe_2023}. The rapid diffusion and adoption of these systems dramatically increases the surface area of potential tail risks and presents difficult governance challenges.

\section{Lessons for AI governance}\label{lessons-for-ai-governance}

Understanding AI systems as complex systems illuminates important governance challenges, many of which are overlooked by current regulatory frameworks \cite{Arbel_Tokson_Lin_2024,Kolt_2024a,Shur-Ofry_2024}. To illustrate, the European Union's AI Act defines ``systemic risk'' from AI as ``actual or \emph{reasonably foreseeable} negative effects on public health, safety, public security, fundamental rights, or the society as a whole'' \cite{noauthor_2024} (Art. 3(65))---an approach that largely ignores the unpredictable and cascading nature of risks in complex interconnected systems. Employing a complexity perspective allows us to draw on regulatory insights regarding complex systems in other domains, including climate policy \cite{Cosens_Ruhl_Soininen_Gunderson_Belinskij_Blenckner_Camacho_Chaffin_Craig_Doremus_Glicksman_Heiskanen_Larson_Similä_2021,Craig_2010,Ruhl_2010}, financial regulation \cite{Arner_2010,Schwarcz_2008, Schwarcz_2009}, and public health \cite{Malcai_Shur-Ofry_2021}, and offer guiding principles for tackling the governance challenges posed by AI.

\subsection{Regulating under deep uncertainty}\label{regulating-under-deep-uncertainty}

While the regulation of any moving target is difficult \cite{Collingridge_1980,Moses_2007, noauthor_2011,Crootof_Ard_2020}, the regulation of AI systems characterized by rapid development, emergent properties, feedback loops, and unpredictable cascading effects is a particularly thorny problem \cite{Arbel_Tokson_Lin_2024,Kaminski_2023,Kolt_2024a}. Neither technologists nor policymakers can reliably predict the capabilities of AI systems or accurately forecast their negative externalities \cite{Ganguli_Hernandez_Lovitt_Askell_Bai_Chen_Conerly_Dassarma_Drain_Elhage_ElShowk_Fort_Hatfield-Dodds_Henighan_Johnston_Jones_Joseph_Kernian_Kravec_Mann_Nanda_Ndousse_Olsson_Amodei_Brown_Kaplan_McCandlish_Olah_Amodei_Clark_2022,Kolt_2024a}. Regulatory efforts, whether targeted at model development or deployed systems and applications, must contend with an ongoing information deficit \cite{Doremus_2010,Karkkainen_2000, Karkkainen_2007} and deep uncertainty \cite{Kay_King_2020, noauthor_2019}. The problem, at is core, is that by the time the capabilities and real-world ramifications of AI systems are properly understood, it may be too late to intervene effectively---a challenge familiar to policymakers in other domains \cite{Collingridge_1980}.

In light of these challenges, we propose three desiderata for designing AI governance mechanisms: (1) policymakers should have the capacity and resources to take early and scalable regulatory action; (2) regulatory action should be dynamic and highly responsive to changing conditions; and (3) policymakers should adopt complexity-compatible risk thresholds with respect to AI systems that exhibit properties characteristic of complex systems.

\subsubsection{(1) \emph{Early and scalable intervention}}\label{early-and-scalable-intervention}

When risks cascade in complex systems, policymakers must be able to respond early, rapidly, and at scale \cite{Cohen_Kolt_Bengio_Hadfield_Russell_2024,Kolt_2024a}. For example, to prevent large-scale economic harm from vulnerabilities in a widely used automated stock trading tool, regulators may need to intervene before the harm has (fully) materialized. Counterintuitively, the case for robust intervention to govern complex systems, including AI technologies, may decline over time \cite{Malcai_Shur-Ofry_2021}. While early intervention (made on the basis of only limited information) could prevent the relevant harm, intervention at a later point (made on the basis of more complete information) may in fact no longer be effective \cite{Malcai_Shur-Ofry_2021}. By analogy, lockdowns and border closures designed to prevent the spread of a pandemic are far more effective, and hence more justifiable, earlier in time (despite the absence of complete information), before the pandemic has spread beyond the ability to contain it, after which such mandates may no longer be as effective \cite{Malcai_Shur-Ofry_2021}. A similar dynamic could apply to AI technologies that exhibit emergent properties, diffuse rapidly and nonlinearly, and lead to cascading effects the could cause large-scale harm.

Apart from the timing of intervention, mechanisms for governing AI systems must also operate at sufficient scale \cite{Ford_2013,Kolt_2024a,Ruhl_2010}. Continuing with the example of vulnerabilities in a widely used automated stock trading tool, for governance mechanisms to be effective they will need to operate successfully across a very large number of actors, institutions, and environments that interact with the tool in question. Consequently, certain conventional governance mechanisms such as manual human oversight and evaluation may be ineffective \cite{Crootof_Kaminski_Price_2023}, while more scalable mechanisms such as automated oversight and evaluation may, despite their shortcomings and potential risks, become necessary \cite{Bowman_Hyun_Perez_Chen_Pettit_Heiner_Lukošiūtė_Askell_Jones_Chen_Goldie_Mirhoseini_McKinnon_Olah_Amodei_Amodei_Drain_Li_Tran-Johnson_Kernion_Kerr_Mueller_Ladish_Landau_Ndousse_Lovitt_Elhage_Schiefer_Joseph_Mercado_DasSarma_Larson_McCandlish_Kundu_Johnston_Kravec_Showk_Fort_Telleen-Lawton_Brown_Henighan_Hume_Bai_Hatfield-Dodds_Mann_Kaplan_2022,Perez_Huang_Song_Cai_Ring_Aslanides_Glaese_McAleese_Irving_2022,Perez_Ringer_Lukosiute_Nguyen_Chen_Heiner_Pettit_Olsson_Kundu_Kadavath_Jones_Chen_Mann_Israel_Seethor_McKinnon_Olah_Yan_Amodei_Amodei_Drain_Li_Tran-Johnson_Khundadze_Kernion_Landis_Kerr_Mueller_Hyun_Landau_Ndousse_Goldberg_Lovitt_Lucas_Sellitto_Zhang_Kingsland_Elhage_Joseph_Mercado_DasSarma_Rausch_Larson_McCandlish_Johnston_Kravec_ElShowk_Lanham_Telleen-Lawton_Brown_Henighan_Hume_Bai_Hatfield-Dodds_Clark_Bowman_Askell_Grosse_Hernandez_Ganguli_Hubinger_Schiefer_Kaplan_2023}.

\begin{table}[hb!]
\small\renewcommand{\arraystretch}{1.5}
\caption{Adaptation mechanisms in prominent governance frameworks}
\begin{tabular}{@{\hspace{-5pt}}>{\centering}p{0.23\linewidth}>{\centering}p{0.23\linewidth}
p{0.45\linewidth}@{}}
\rowcolor{gray!60}\bfseries\Tstrut  & \bfseries\Tstrut\normalsize Type of \mbox{framework}\rule[-2.5mm]{0pt}{4mm} & \bfseries\Tstrut\centering\normalsize\arraybackslash \parbox{\linewidth}{\centering Adaptation}  \parbox{\linewidth}{\centering mechanisms} \\
\rowcolor{gray!05} {\textbf{U.S. National Institute of Standards and Technology AI Risk Management Framework}} (January 2023) \cite{noauthor_2023a} & Non-binding practice and policy framework & Describes the framework as a ``living document'' and refers to current document as v 1.0. Stipulates that NIST will regularly review the document, including with formal public input. \\
\rowcolor{background} {\textbf{China Interim Measures for the Management of Generative Artificial Intelligence Services}} (August 2023) \cite{PRC_2023} & Binding obligations on generative AI service providers & These interim measures are likely to be superseded by the draft Artificial Intelligence Law of the People’s Republic of China first circulated in March 2024 \cite{PRC_2024}. \\
\rowcolor{gray!05}{\textbf{U.S. Executive Order on the Safe, Secure, and Trustworthy Development and Use of Artificial Intelligence}} (October 2023) \cite{TheWhiteHouse_2023} & Binding reporting requirements and mandatory government actions & Requirements carried out by various government agencies that exercise significant discretion. Like other US executive orders, the order can be modified or revoked by the President---and was revoked by President Trump in January 2025 \cite{TheWhiteHouse_2025}. \\
\rowcolor{background}{\textbf{European Union Artificial Intelligence Act}} (August 2024) \cite{noauthor_2024} & Binding cross-sector regulation and establishment of new regulatory institutions & While the Act itself will be difficult to amend, it includes mechanisms for amending certain key provisions and further changes through implementing acts, delegated acts, externally determined standards, and a code of practice for general-purpose AI \cite{GPAI_2025}. \\
\end{tabular}%
\end{table}

\subsubsection{(2) \emph{Adaptive governance}}\label{adaptive-governance}

Even if the above desideratum is met, governance institutions will nonetheless need to adapt to new conditions arising due to hard-to-predict changes in AI systems, their usage, and the broader sociotechnical context in which they operate \cite{Benthall_Sivan-Sevilla_2024,Reuel_Undheim_2024}. To this end, policymakers should draw on the principles of adaptive management and resilience proposed in the field of climate policy and environmental governance \cite{Folke_Hahn_Olsson_Norberg_2005,Holling_1973, noauthor_1978}. According to these principles, governance mechanisms that aim to regulate complex systems should not be static institutions, but feedback-driven processes that iteratively respond and adapt to new information while preserving overarching societal goals and values \cite{Doremus_2010,Ruhl_2005, Ruhl_2010}. This dynamic approach to governance is especially crucial for mitigating cascading failures of complex systems \cite{DSouza_2017,Ruhl_2019}. 

As illustrated in Table 1, prominent AI governance frameworks include mechanisms for adaptation and change. Notably, these mechanisms for adaptation do not specify the type of information required to bring about changes in the corresponding governance framework. For example, it is unclear what information the EU would need to receive in order to add or remove AI systems or applications from the list of high-risks systems in the EU AI Act \cite{noauthor_2024}. That being said, this feature of regulatory frameworks is not necessarily a defect. Perspectives from complex systems suggest that overly fine-grained rules that attempt to anticipate every possible contingency are inherently limited \cite{Vivo_Katz_Ruhl_2024}. Accordingly, the use of open-ended standards in AI regulation that accommodate regulatory discretion and responsiveness---without stipulating the precise type of information required to trigger regulatory action---have notable advantages and may, on balance, be preferable to more prescriptive approaches.

Nevertheless, to be effective, adaptive governance must be guided by up-to-date information concerning the systems being governed \cite{Coglianese_Zeckhauser_Parson_2004,Doremus_2010,Karkkainen_2000, Karkkainen_2007,Stephenson_2010,VanLoo_2019}. In the case of AI, policymakers must continually acquire information about the current and anticipated capabilities, trends, and impacts of AI systems \cite{Whittlestone_Clark_2021,Clark_2023,Kolt_Anderljung_Barnhart_Brass_Esvelt_Hadfield_Heim_Rodriguez_Sandbrink_Woodside_2024,Reuel_Bucknall_Casper_Fist_Soder_Aarne_Hammond_Ibrahim_Chan_Wills_Anderljung_Garfinkel_Heim_Trask_Mukobi_Schaeffer_Baker_Hooker_Solaiman_Luccioni_Rajkumar_Moës_Ladish_Guha_Newman_Bengio_South_Pentland_Koyejo_Kochenderfer_Trager_2024}. In other words, they must engage in ``evidence-seeking'' policy \cite{casper2025pitfalls}. Adaptive governance also implies that policymakers should be cognizant of potential abrupt changes in a system's performance or behavior, and of the possibility that such changes will quickly percolate and affect interconnected systems. 

Current regulatory frameworks have made significant progress in tackling this information problem, establishing multiple mechanisms for furnishing policymakers with decision-relevant information. For instance, the EU AI Act requires companies to keep detailed records of certain ``high-risk'' AI systems and proposes mechanisms for monitoring these systems and reporting safety incidents \cite{noauthor_2024} (Arts. 11--12, 72--73). Notwithstanding these mechanisms, deciding how to interpret the information gathered, and whether (or how) to act upon it, still presents a significant challenge for policymakers.

\subsubsection{(3) \emph{Complexity-compatible risk thresholds}}\label{complexity-compatible-risk-thresholds}

What threshold of risk from AI should trigger regulatory intervention? \cite{Koessler_Schuett_Anderljung_2024} What information would constitute sufficient evidence that such a threshold has been reached? Regulators often dodge these questions by postponing governance decisions until the relevant ``evidentiary burden'' is satisfied \cite{Galle_2015}. The problem with this approach is that, because many AI systems exhibit properties characteristic of complex systems, such information may only become available at a time after which intervention has become more costly or less effective \cite{Collingridge_1980,Malcai_Shur-Ofry_2021,Posner_2004,Sunstein_2005, noauthor_2011}. 

Consequently, to intervene effectively policymakers may need to relax the policy-relevant informational threshold and resort to ``satisficing'' \cite{Simon_1955}---i.e., making governance decisions on the basis of incomplete information collected at an earlier stage in the technology's development and use \cite{Malcai_Shur-Ofry_2021}. For example, policymakers may need to amend AI safety standards upon receiving interim red-teaming results that indicate certain dangerous capabilities prior to receiving the final results, let alone comprehensive studies establishing the precise probability or magnitude of the relevant risks. In such cases, rather than wait until more detailed or complete information is available, regulators should familiarize themselves with the patterns characteristic of complex systems in order to evaluate the potential risks from AI systems and design appropriate interventions.

One potential route is to employ a legal doctrine known as the ``precautionary principle''. The principle, which is used in environmental governance and public health policy, supports preemptive regulatory intervention before harms have (fully) materialized or risks are established conclusively, often requiring actors interested in pursuing a potentially risky activity to first prove its safety \cite{Harremoes_Gee_MacGarvin_Stirling_Keys_Wynne_Vaz_2013,Nash_2008}. A prominent criticism of the precautionary principle---which in the case of AI may require robust technical safety guarantees \cite{Tegmark_Omohundro_2023,Dalrymple_Skalse_Bengio_Russell_Tegmark_Seshia_Omohundro_Szegedy_Goldhaber_Ammann_Abate_Halpern_Barrett_Zhao_Zhi-Xuan_Wing_Tenenbaum_2024} or other forms of assurance \cite{Buhl_Sett_Koessler_Schuett_Anderljung_2024,Clymer_Gabrieli_Krueger_Larsen_2024} ---is that it does not withstand cost-benefit analysis, i.e., it unduly limits or forgoes the gains from new technology \cite{Farber_2010,Posner_2004,Sunstein_2002}. 

However, as explored in the context of pandemic responses, insights from complexity can help refine and calibrate the precautionary principle. In particular, regulators should consider whether the relevant risks will likely spread swiftly and exponentially and thereby pose grave systemic risk \cite{Malcai_Shur-Ofry_2021}. Where this is the case, the costs of postponing regulatory intervention until more complete information is obtained are often multiplicative, such that delay can be orders of magnitude costlier than early intervention. For example, refraining from intervening to prevent failures in an AI system connected to critical infrastructure could result in costly damage that rapidly percolates into other safety-critical systems \cite{Buldyrev_Parshani_Paul_Stanley_Havlin_2010,Gao_Bashan_Shekhtman_Havlin_2022,Li_Bashan_Buldyrev_Stanley_Havlin_2012,Yang_Nishikawa_Motter_2017, DSouza_2017, Ruhl_2019}. Conversely, the costs of early intervention (e.g., requiring additional guardrails in response to interim red-teaming results) are often linear and additive. Seen through the lens of complexity, cost-benefit analysis can in certain cases support a precautionary approach to governing AI. A central consideration in this analysis is the level of interconnectedness between AI systems and other sociotechnical systems \cite{Shur-Ofry_2024}, as well as the potential for feedback loops and cascading effects. Future work will need to examine these considerations in specific settings in order to implement complexity-compatible risk thresholds in practice.

\section{Outlook}\label{outlook}

The hallmarks of complex adaptive systems increasingly exhibited by AI systems---nonlinearity, emergence, feedback loops, cascading effects, and tail risks---underscore the difficult governance challenges facing policymakers. Studying AI through the lens of complexity can guide policymakers to focus on the well-studied patterns of complex systems that are likely to arise in AI systems. Complexity theory helps identify and characterize new risks from AI systems, and points toward more appropriate governance mechanisms. Policymakers addressing the challenges from AI should draw on approaches developed in other domains that confront complexity-related challenges, including climate policy and public health. As AI systems continue to advance and diffuse, the time is ripe to deepen these interdisciplinary connections.

\newpage
\printbibliography

\end{document}